\documentclass[letterpaper]{jpconf}

\usepackage{bm}
\usepackage{graphicx}
\usepackage{natbib}

\begin{document}
\title{Quantum chromodynamics with advanced computing}

\author{Andreas S Kronfeld}

\address{Fermi National Accelerator Laboratory,
	Batavia, IL 60510, USA\\[0.7em]
	(for the USQCD Collaboration)}

\ead{ask@fnal.gov}

\begin{abstract}
We survey results in lattice quantum chromodynamics from groups in the 
USQCD Collaboration.
The main focus is on physics, but many aspects of the discussion are 
aimed at an audience of computational physicists.
\end{abstract}

\section{Introduction and background}

Quantum chromodynamics (QCD) is the modern theory of the strong nuclear 
force.
The physical degrees of freedom are quarks and gluons.
The latter are the quanta of gauge fields, in some ways analogous to
photons.
The crucial difference is that gluons couple directly to each other, 
whereas photons do not.
A~consequence of the self-coupling is that the force between quarks
does not vanish at large distances (as the Coulomb force $e^2/(4\pi r^2)$
does) but becomes a constant $F_{\rm QCD}\approx%
800~\textrm{MeV}\,\textrm{fm}^{-1}\approx 15,000~\textrm{N}$.
It would, thus, require a vast amount energy to separate quarks out to
a macroscopic distance.
Indeed, long before the separation becomes large enough to measure,
the energy stored in the gluon field ``sparks'' into quark-antiquark
pairs.
This phenomenon of QCD explains why freely propagating quarks are not
observed in nature, and it is called confinement.

QCD is a quantum field theory, which means that from the outset one 
must deal with mathematical objects that are unfamiliar even to many 
physicists.
The most widely used theoretical tool for quantum field theories is
relativistic perturbation theory, as developed for quantum
electrodynamics (QED) by Feynman, Schwinger, Tomonaga and others sixty
years ago~\cite{Schwinger:QED}.
Perturbation theory provides the key connection between the
mathematical theory with experiments in QED~\cite{Kinoshita:1996vz}
and the Glashow-Weinberg-Salam theory merging QED with the
weak nuclear force~\cite{Alcaraz:2007ri}.

\begin{figure}[bp]
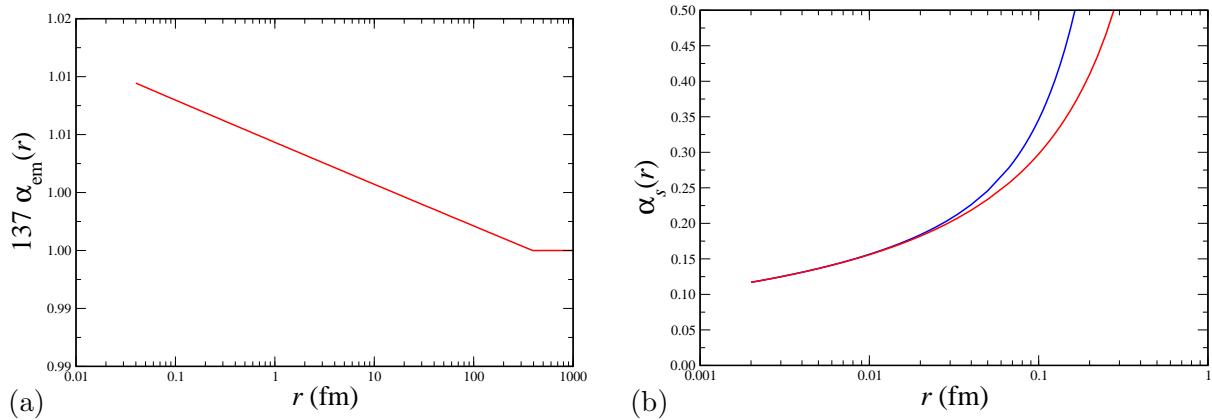

	(a)\hspace*{-1.2em}\includegraphics[width=0.48\textwidth]{run_qed}\hfill
	(b)\hspace*{-1.2em}\includegraphics[width=0.48\textwidth]{run_r}
	\caption[fig:run]{Running of (a) the QED and (b) the QCD coupling
	with distance.
	For QCD the red curve shows the perturbative result, and 
	the (higher) blue show the effect of confining forces.}
	\label{fig:run}
\end{figure}

A textbook example of perturbative quantum field theory is to calculate 
how virtual pairs induce a distance dependence on the coupling in 
quantum gauge theories.
In QED, one considers the fine structure 
constant,
\begin{equation}
	\alpha = \frac{e^2}{4\pi\hbar c},
	\label{eq:alpha}
\end{equation}
where $-e$ is the charge of the electron.
In QCD, one has the strong coupling~$\alpha_s$, related to the gauge
coupling~$g$ by analogy with equation~(\ref{eq:alpha}).
When virtual pairs are taken into account, $\alpha$ and $\alpha_s$ are 
not constant but depend on distance; they are said to ``run.''
The running, depicted in figure~\ref{fig:run}, is completely different
in the two theories.
In QED $\alpha$ becomes constant at distances $r>\hbar/m_ec$ (the 
electron's Compton wavelength), but at short distances it grows.
In QCD $\alpha_s$ grows at long distance; perturbative running no longer
makes sense once $r$ is in the confining regime.
At short distances---or, equivalently, high energies---the QCD 
coupling is small enough that perturbation theory can again be used.
For example, the cross section for an electron-positron pair to
annihilate and produce a quark-antiquark pair can be reliably computed
as a power series in $\alpha_s$.
The quark and antiquark each fragment into a jet of hadrons---mostly
pions, but some protons and neutrons, too---and general properties such
as the energy flow and angular distributions of these jets can be
traced back to the quark and antiquark.
In this way, the experiments can measure quark properties, and these 
are found to agree with theoretical calculations~\cite{Dixon:2007hh}.

QCD and the electroweak theory form the foundation of the Standard 
Model of elementary particles.
They are so well established that many particle physicists consider the
chromodynamic and electroweak gauge symmetries, and the associated
quantum-number assignments of the quarks and leptons, to be laws of
nature.
We cannot conceive of a more comprehensive theory that does not
encompass them.
But there is more to the Standard Model.
We know that something must spontaneously break the electroweak
symmetry and that something (perhaps the same thing) must generate
masses for the quarks and leptons.
The Standard Model contains interactions that \emph{model} these
phenomena.
The model interactions are consistent with observations yet also
incomplete.
It is hoped that insights obtained with the Large Hadron Collider
(LHC), to commence operations at CERN later this year, will enable
particle physicists to solve many of the puzzles raised by the
Standard Model.

The purpose of this paper is to cover QCD, particularly in the 
strong-coupling regime where perturbation theory is insufficient.
It is worth bearing in mind, however, that there is a close connection
between many of the results discussed below and some of the big
questions in particle physics.
A~simple example is the masses of the quarks.
To determine the quark masses (which because of confinement cannot be
measured), we have to calculate the relation between quark masses and 
measurable quantities, such as hadron masses.
Quark masses are interesting for many reasons, the most unsettling of
which is as follows.
\begin{figure}[bp]
\fbox{\begin{minipage}{0.4\textwidth}
	Quark flavors in the Standard Model:
	$$ \begin{array}{ccc}
	\left(\begin{array}{c} u \\ d \end{array} \right) &
	\left(\begin{array}{c} c \\ s \end{array} \right) &
	\left(\begin{array}{c} t \\ b \end{array} \right) \\[1em]
	m_u < m_d & m_c > m_s & m_t > m_b \\
	!!! & & 
	\end{array} $$
	What explains the pattern of masses?
\end{minipage}}
\hfill
\begin{minipage}{0.5\textwidth}
	\caption{The six \emph{flavors} of quarks come the three 
	\emph{generations} with a doublet in each generation.
	The doublet structure plays a role in the electroweak interactions.
	For example, the upper (lower) entries have electric charge $+2/3$ 
	($-1/3)$.}
	\label{fig:mqineq}
\end{minipage}
\end{figure}
The mass of the top quark is larger than that of the bottom quark;
see figure~\ref{fig:mqineq}.
Similarly, the mass of the charmed quark is larger than that of the 
strange quark.
If this simple pattern held with the up and down quarks---the ones that
make up proton and neutrons---then the up quark would be more massive
than the down.
But then protons would decay to neutrons, positrons, and neutrinos 
($p\to ne^+\nu$).
The positrons would find electrons and annihilate to photons.

This universe would consist of neutron stars surrounded by a swarm
photons and neutrinos, and nothing else.
Our universe is not at all like this, because neutrons are more
massive than protons.
The allowed decay reaction is $n\to pe^-\bar{\nu}$, leading to an
abundance of protons and electrons, and making possible chemistry and
biochemistry.
With QCD we trace the neutron-proton mass difference back to the 
down-up quark masses and then to the deeper origin of quark masses.
The pattern of quark masses is necessarily not simple, and the search 
for simple explanation of a messy pattern occupies many particle 
theorists.
Below we give further examples where QCD is essential for 
working out the mysteries of particle physics.

QCD is also the cornerstone of modern nuclear physics.
The simplest nucleus is nothing but the proton, a bound state of two 
up quarks and one down quark.
The neutron is a bound state of two down quarks and one up quark.
The confining force is very, very strong.
By comparison, the force traditionally called the strong nuclear 
force is a residue of the fundamental chromodynamic force.
This is similar to van~der~Waals forces among molecules, which are 
electromagnetic forces between neutral objects with structure leading 
to a distribution of electric charge.
Some basic problems are, therefore, to understand nucleon structure 
directly from QCD, to study the excitation spectrum of nucleons 
(and their cousins with strangeness and other flavors), and 
to see how few-nucleon (or, more generally, few-hadron) systems 
interact.

Another active area of research in nuclear QCD is the properties of
quark matter at high temperature and density.
This regime is important to cosmology, because an epoch of the early
universe consisted of hot quark matter, and to astrophysics, because,
for example, neutron stars are composed of dense quark matter.
In the laboratory, nuclear physicists create hot, dense quark matter
by colliding heavy ions, at present with Brookhaven's RHIC, soon with
(the heavy-ion mode of) CERN's LHC, and someday with GSI's~FAIR.

At strong coupling it is necessary to go beyond perturbation theory.
In a relativistic quantum field theory, such as QCD, this is not easy.
An important obstacle is that a quantum field associates one (or 
perhaps a few) degrees of freedom with every point in space-time.
Hence, the number of degrees of freedom is uncountably infinite.
Furthermore, the fluctuations at short distances, if added up naively,
contribute an ultraviolet divergence to anything of interest.
These fluctuations are treated by a set of ideas known as 
\emph{renormalization}~\cite{Lepage:1989hf}.
In short, one must find a tool that (correctly) handles both strongly
coupled fields and the renormalization of short-distance fluctuations.

This tool is lattice field theory or, for gauge theories like QCD, 
lattice gauge theory.
The idea, which goes back at least as far as Heisenberg, is to replace
continuous space (or, in practice, space-time) with a discrete grid, or
lattice.
Then the number of degrees of freedom is countable.
If the lattice is of finite extent, it occupies a finite physical
volume, and, moreover, the number of degrees of freedom is finite.
In 1974, Kenneth Wilson showed how to maintain gauge symmetry with a 
discrete lattice~\cite{Wilson:1974sk}.
In this and subsequent work, he and others elaborated how the lattice
provides a mathematically rigorous definition of quantum field theory
via functional integrals~\cite{Glimm:1987ng}.

The functional integral for quantum fields on a finite lattice is a
familiar object, a multi-dimensional integral, that can be evaluated on
a computer.
Before downloading some publicly available code~\cite{usqcd} and trying
to run it on a laptop, let us pause to contemplate how many variables of
integration are reasonable (for physics).
One wants to have the spacing between sites on the lattice small 
enough to resolve hadron structure, and one wants the volume large 
enough that the boundary conditions do not modify the structure.
If one supposes that the ratio of these two lengths is 32, then the
dimensions of the integrals of interest are
\begin{equation}
	{\tt gluon~d.o.f.} = 8\times 4\times 32^3 \times 128 > 10^8,
\end{equation}
because gluons come in 8~colors with 4 polarization states each.
The factor 128 presumes a space-time lattice with a time extent long (in
relativistic units) compared to the spatial extent; this is common
practice.

To cope with integrals of such large dimension, the only practical 
technique is Monte Carlo integration with importance sampling.
The weight guiding the importance sampling is $e^{-S}$, where $S$ is 
the classical action for the random sample of gluon variables.
In functional integral formalisms for quantum mechanics, the 
expression for the action is the defining equation of the physical 
system.
The development of algorithms to generate these samples is a vibrant
subject, covered in part by B\'alint Jo\'o's poster here at the SciDAC
conference.
This paper will not dwell on algorithms, but it is important to 
mention one more complication.

Quarks are fermions and, as such, must satisfy the Pauli exclusion
principle.
In the functional integral formalism, this is handled by introducing
anticommuting Grassmann variables for fermions.
The integration is no longer Riemann (or Lebesgue) integration but a
formal procedure called Berezin integration.
To make a long story short, in lattice QCD we always can and do carry 
out the Berezin integration by hand.
The outcome is that the weight for the importance sampling becomes 
$\det M \exp(-S_{\rm gluons})$, where $M$ is a sparse matrix with 
space-time indices.
The matrix $M$ is $N\times N$, where 
\begin{equation}
	N = {\tt quark~d.o.f.} = n_f \times 3\times
		2\times 2\times 32^3 \times 128 > 10^8.
\end{equation}
The quark field represents $n_f$ flavors with 3~colors and 2~spin states of
quark and antiquark for every flavor and color.
The matrix $M$ is sparse because it is a lattice version of a 
differential operator, the Dirac operator (generalized to QCD).
Incorporating $\det M$ into the weight is computationally demanding, 
increasing by two or three orders of magnitude the amount of
floating-point operations needed to generate a new configuration of 
gluons.

The lattice is a helpful device mathematically and computationally, 
but it is not physical.
To obtain a real result of the chromodynamics of continuum spacetime, 
one must work out the integrals for a sequence of lattices, and take 
the limit $a\to 0$ in a way that respects renormalization.
Taking $a$ smaller and smaller increases the computing burden as
onerously as $a^{-(4+z)}$.
The exponent $z$ depends on the algorithm and is typically around 1 or~2.
The 4 in the exponent is the key reason why numerical lattice QCD
requires the most advanced computing facilities available.
No algorithm can reduce it, because it is the dimension of space-time.

Most lattice-QCD calculations are carried out in two steps.
The first is to generate an ensemble of gluon fields, with a certain 
lattice spacing and quark masses.
This requires the computationally super-demanding $\det M$.
To generate an ensemble of useful size, one needs several hundred, 
perhaps even a few thousand, samples of the gluon field.
Computers of the highest available capability are needed for this step.
In recent years, lattice gauge theorists have carried out this step 
with the special-purpose computer QCDOC, large clusters of PCs, 
and, more recently, leadership-class machines constructed under the 
auspices of the DOE's Office of Advanced Scientific Computing 
Research (ASCR).
These ensembles of gluon fields are valuable: within the U.S.\ they are 
generated by and for particle and nuclear physicists and shared under 
various agreements.
File-sharing is facilitated by formats and software developed by the 
International Lattice Data Grid~\cite{DeTar:2007au}.

The second step is to mine these ensembles for physics.
A simple example is to compute the quantum mechanical amplitude for a
proton to propagate from one point to another, and study the behavior of
the amplitude as the separation varies: this is what one does to compute
the proton mass, in fact.
The key ingredients are a few rows of the inverse of the matrix~$M$,
introduced above.
With the same ensemble one can study protons, pions, and more uncommon
hadrons such as the charmed strange pseudoscalar meson (or $D_s$ for
short).
Different rows of~$M^{-1}$ (i.e., different flavors, colors,
spins, and space-time separations) are combined in various ways to
obtain the quantum numbers of the hadrons in the problem at hand.
Because each problem is different, and there are so many of them, we
must attack this problem with the highest capacity computers.
The job mix here is heterogeneous, with many users whose job streams
range from many large jobs to extremely many medium-sized jobs.
The challenge is to develop systems that scale well (for the large 
jobs), while maintaining the flexibility needed to handle the
heterogeneity.
The USQCD Collaboration has designed clusters of PCs that deliver 
this capacity in a cost-effective way~\cite{Holmgren:2004nk}.

The remainder of this paper focuses on physics.
It is organized according to four principal themes of lattice gauge
theory:
\begin{itemize}
	\item Determination of fundamental parameters of the Standard 
	Model of particle physics
	\item Study of nucleon structure, the hadron spectrum, and hadron 
	interactions
	\item Simulation of the thermodynamic properties of QCD at 
	nonzero temperature and density
	\item Lattice gauge theories beyond QCD, such as those appearing 
	in models of electroweak symmetry breaking
\end{itemize}
Roughly speaking, the first and last are particle physics, 
and the second and third are nuclear physics.
But the lines are blurry: for example, some of the simplest calculations of
proton structure (the second item) may play a role more accurate
calculations of $pp$ scattering cross sections at the LHC.
The way that scientific goals drive the computing needs of lattice 
gauge theory are discussed further, along the lines of these themes, in 
public whitepapers of the USQCD Collaboration~\cite{usqcd}.

\section{Standard Model particle physics}
\label{sec:sm}

In quantum electrodynamics, the fundamental parameters are the masses of
electrically charged particles and the fine structure constant $\alpha$.
Similarly, in quantum \emph{chromo}dynamics, the fundamental parameters
are the masses of colored particles (the quarks) and the strong
coupling~$\alpha_s$.
Just as $\alpha$ can be determined in many ways~\cite{Kinoshita:1996vz},
$\alpha_s$ can be determined at high energies from jet cross sections
with perturbative QCD, and at low energies from the hadron spectrum with
lattice QCD.
Two recent results are (evaluated at distance $r=\hbar/m_{Z^0}c$)
\begin{equation}
	\alpha_s = \left\{
		\begin{array}{llll}
			0.1172 \pm 0.0022 & \textrm{perturbative QCD} & 
				\textrm{jet shapes} & \textrm{Ref.~\cite{Becher:2008cf}} \\
			0.1170 \pm 0.0012 & \textrm{lattice QCD} & 
				\textrm{hadrons} & \textrm{Ref.~\cite{Mason:2005zx}}
		\end{array} \right. ,
\end{equation}
where both error bars reflect experimental and 
theoretical uncertainties.
The first result is determined at high energies, around 100~GeV,
in the regime where quarks can be treated as nearly free.
The second is determined at low energies, of a few GeV and lower, 
where quarks are confined.
Both are linked soundly to the defining equations of QCD.
The agreement demonstrates the richness and broad validity of QCD and 
should make any enthusiastic scientist say ``Wow!''

A straightforward application of lattice QCD is to work out how 
hadron masses depend on quark masses.
Quark masses are interesting for several reasons.
As fundamental constants of nature, they are interesting in their own 
right.
As discussed in the introduction, the pattern of quark masses is 
puzzling and, hence, a motivation to search for extensions of the 
Standard Model that would contain a simple explanation.

Three flavors of quarks---top, bottom, and charm---have masses large
enough to influence some high-energy scattering and decay processes.
Their masses can be determined with perturbative QCD as well as lattice 
QCD.
(The results agree~\cite{Yao:2006px}.)
The other three quarks---up, down, and strange---have masses so small
that the full bound-state problem must be considered.
Lattice QCD, therefore, provides the best information.
Some recent results are the following~\cite{Mason:2005bj,Bernard:2007ps}: 
\begin{eqnarray}
	m_u = 1.9 & \hspace{-5pt}\pm\hspace{-5pt} & 0.2~\textrm{MeV}/c^2, \\
	m_d = 4.6 & \hspace{-5pt}\pm\hspace{-5pt} & 0.3~\textrm{MeV}/c^2, \\
	m_s = 88  & \hspace{-5pt}\pm\hspace{-5pt} &   5~\textrm{MeV}/c^2.
\end{eqnarray}
One remarkable aspect of these results is that the strange quark's 
mass is somewhat smaller than had been thought on the basis of less 
reliable techniques for attacking strongly-coupled QCD.
This point is, perhaps, of interest mostly to particle physicists, but a
further one is of broader interest.
The up and down masses---the constituents of protons and neutrons and,
thus, everyday matter---are very small, within an order of magnitude
of the electron mass ($0.511~\textrm{MeV}/c^2$).
Only about 1--2\% of the proton mass ($938.3~\textrm{MeV}/c^2$) or
neutron mass ($939.6~\textrm{MeV}/c^2$) is accounted for by the quark
masses.
Consequently, most of their masses---and the mass of everyday
matter---arises from confinement, namely, the binding energy of the
gluons and the (relativistic) kinetic energy of bound quarks.
This is a stunning insight, to be revisited in Section~\ref{sec:bsm}.

The caption of figure~\ref{fig:mqineq} states that the pairings of quarks
in each doublet stem from the electroweak interactions.
That is not the full story.
There is no reason for the eigenstates of mass to be the eigenstates of 
the interaction term with $W$ gauge bosons.
The two are related by a unitary transformation, $V_{\rm CKM}$, namely,
\begin{equation}
	\left(\begin{array}{c} d \\ s \\ b \end{array} \right)_W =
		V_{\rm CKM}
		\left(\begin{array}{c} d \\ s \\ b \end{array} \right)_{\rm mass},
	\quad
	V_{\rm CKM} = \left(\begin{array}{ccc} 
		V_{ud} & V_{us} & V_{ub} \\ 
		V_{cd} & V_{cs} & V_{cb} \\ 
		V_{td} & V_{ts} & V_{tb} \end{array} \right).
	\label{eq:CKM}
\end{equation}
Here the initials CKM stand for Cabibbo-Kobayashi-Maskawa; 
Cabibbo introduced equations~(\ref{eq:CKM}) for two generations, and Kobayashi
and Maskawa were the first to study the ramifications for three
generations.

The left-hand side of the first equation denotes the electroweak basis.
The labels on the elements of $V_{\rm CKM}$ are the weak isopartner
and the mass eigenstate, because the transition amplitude for
$W\to\bar{b}u$, for example, is proportional to $V_{ub}$.
The study of physical processes related to $V_{\rm CKM}$ is called
flavor physics, because the central issues are to understand why there
are several flavors of quarks and what lends them their separate
identities.

As a unitary matrix, $V_{\rm CKM}$ contains complex entries.
Some of the phases can be absorbed into unobservable phases of the quark
wave functions.
For only two generations, one physical parameter remains, and it is real.
For three generations, as in nature, three physical parameters are real
and one phase remains.
The complex couplings imparted by the phase lead to physical processes 
that proceed at different rates for particles and the corresponding 
antiparticles.
Such reactions are needed to explain the abundance of matter and the 
dearth of antimatter in the universe.
Among particle physicists, this phenomenon is called $CP$ violation.
It is clearly intriguing to know whether the $CP$ violation 
of the CKM matrix is enough to account for the matter-antimatter 
asymmetry of the universe.
According to present measurements and theoretical understanding, it is
insufficient, so one would like to know what other interactions
violate~$CP$.

In the Standard Model, there is a scalar boson called the Higgs boson.
Section~\ref{sec:bsm} will explain more about it.
For now let us note that both the quark masses and the CKM
matrix have their origin in matrices of couplings between quark fields
and the Higgs field.
Therefore, the puzzles of $CP$ violation and the pattern of the quark 
masses are tied together by the Standard Model.

To make progress in flavor physics, lattice-QCD calculations are needed 
to interpret the experimental measurements.
In a schematic form the rate $\Gamma$ is given by
\begin{equation}
	\Gamma = \left(
		\begin{array}{c} \textrm{known} \\ \textrm{factor} \end{array}
		\right) \left(
		\begin{array}{c} \textrm{CKM} \\ \textrm{factor} \end{array}
		\right) \left(
		\begin{array}{c} \textrm{QCD} \\ \textrm{factor} \end{array}
		\right) ,
\end{equation}
where the ``known factor'' consists of well-measured physical constants 
and numerical factors like~$4\pi$.
The paradigm is to measure as many flavor-changing processes as 
possible, and use the Standard-Model formulae to over-determine the CKM 
matrix.
If all the determinations are consistent, then one can set limits on
non-Standard sources of flavor and $CP$ violation.
To do so, the QCD factor must be computed, and often the only way to do 
so is with lattice QCD.

\begin{figure}[tbp]
	\centering
	\includegraphics[width=\textwidth]{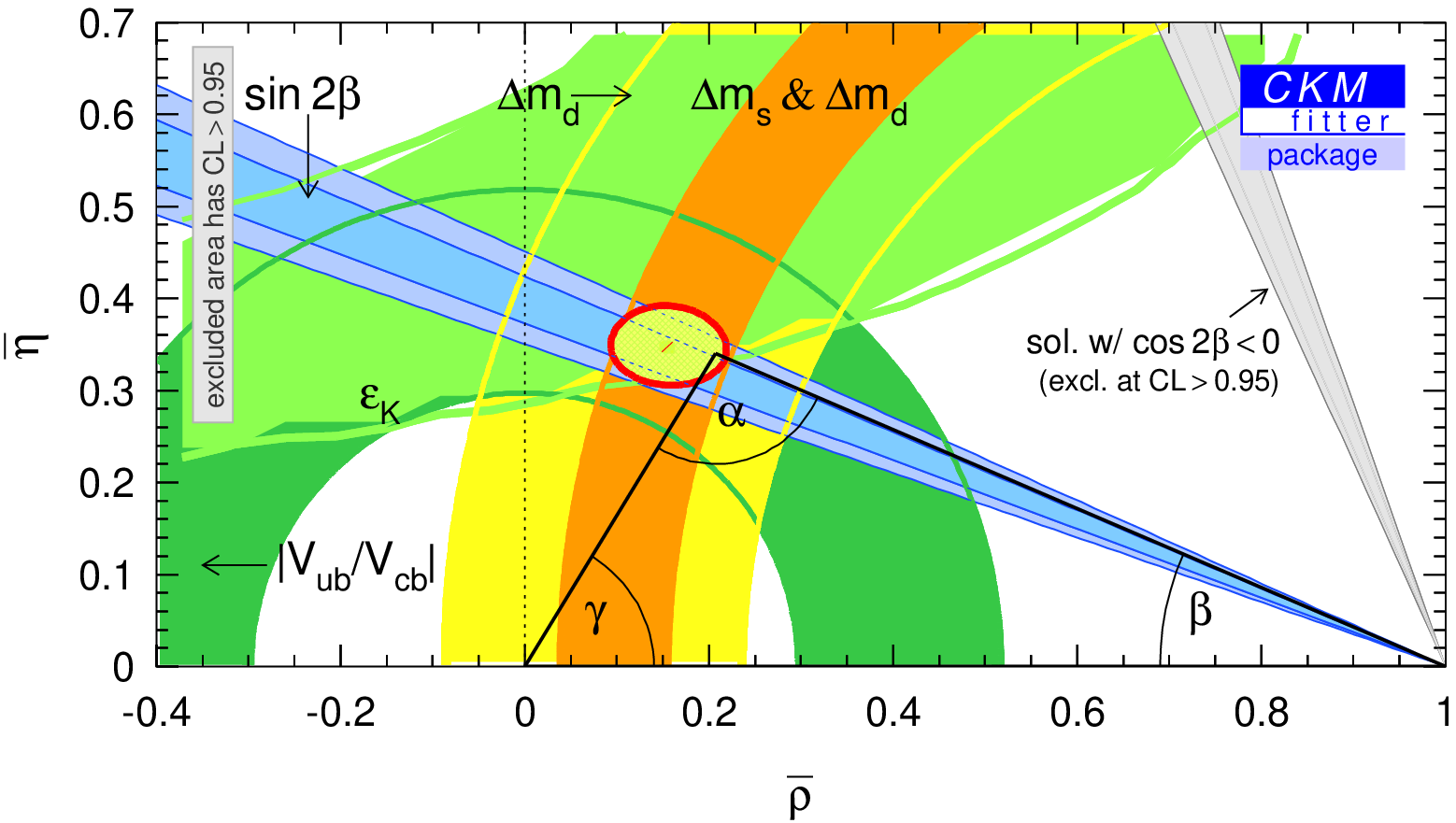} \\
	\includegraphics[width=\textwidth]{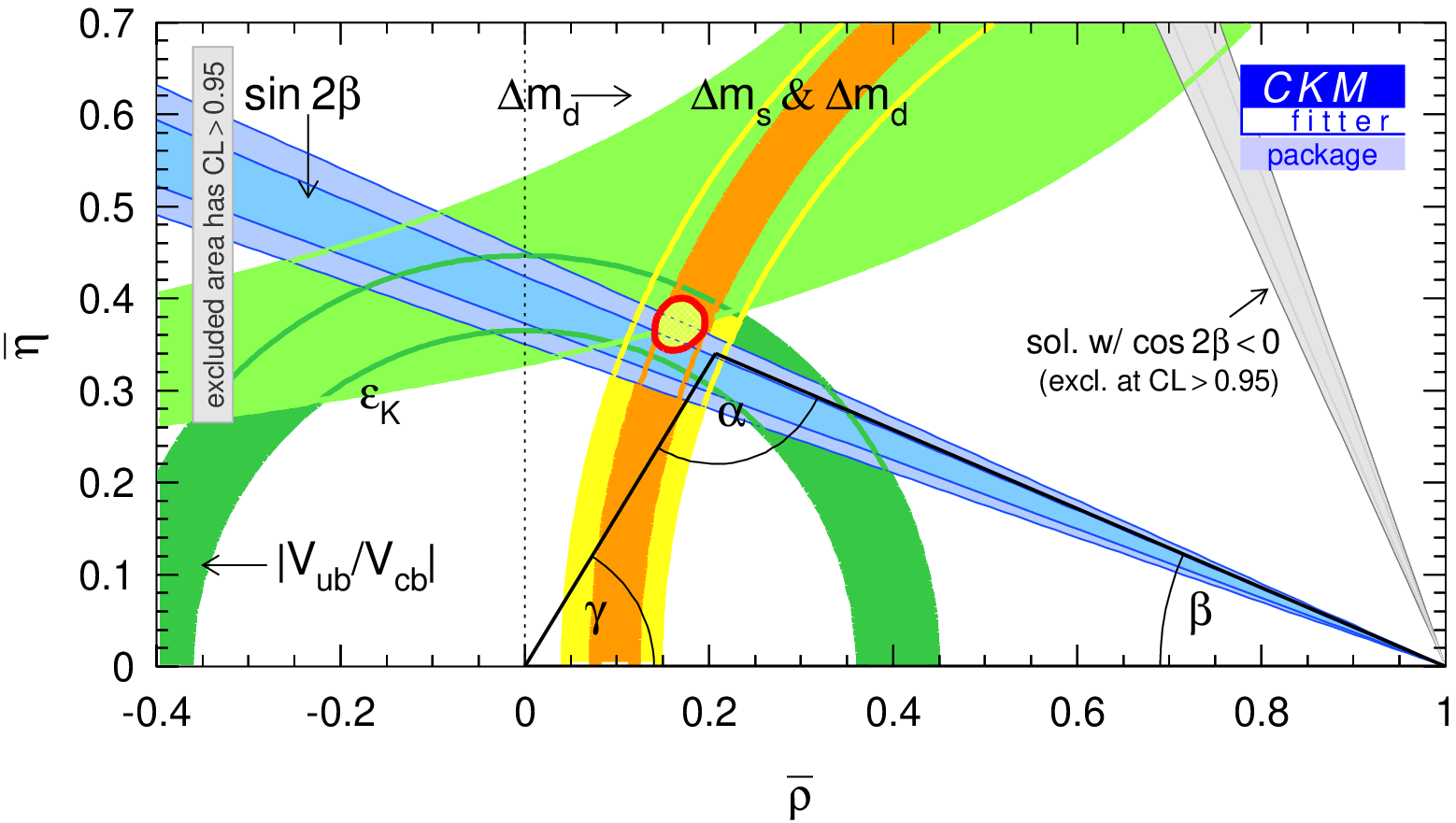}
	\caption[fig:ut]{Constraints on the CKM unitarity triangle from 
	experiments and QCD calculations, highlighting those where lattice 
	QCD plays a role.
	The upper panel shows the current status, and the lower a forecast
	of how the uncertainties can be reduced with lattice
	QCD~\cite{VandeWater:2007zz}.
	In the lower panel, one can envision tension between the lime-green 
	hyperbola and the blue wedge, depending on how the central values 
	evolve and measurements and lattice-QCD computations improve.
	The current status of the angles $\alpha$ and $\gamma$ is omitted, 
	because better measurements are needed; they are expected from LHC$b$.}
	\label{fig:ut}
\end{figure}

There are dozens of relevant measurements, but because the CKM matrix
is unitary, the elements of the CKM matrix must satisfy constraints.
A particular vivid one is the unitarity triangle that stems from the
orthogonality of columns of a unitary matrix:
\begin{equation}
	V_{ud}^* V_{ub} + V_{cd}^* V_{cb} + V_{td}^* V_{tb} = 0,
	\label{eq:ut}
\end{equation}
which traces out a triangle in the complex plane.
The triangle and the present uncertainties from various constraints is 
shown in figure~\ref{fig:ut}, dividing each side by $V_{cd}^*V_{cb}$.
Except for the wedge labelled $\sin\beta$
the uncertainties can be reduced with lattice 
calculations of the QCD factor.

Lattice calculations for a wide range of observables relevant to the CKM
matrix have appeared over the past few years, and further improvements
are under way.
These include calculations of properties of strange
kaons~\cite{Bernard:2007ps,Aubin:2004fs,Beane:2006kx,%
Gamiz:2006sq,Antonio:2007pb,Boyle:2007qe},
charmed $D$ and $D_s$ mesons~\cite{Aubin:2004ej,Aubin:2005ar,Follana:2007uv},
and beautiful $B$ and $B_s$ mesons~\cite{Okamoto:2004xg,Dalgic:2006dt,%
Gray:2005ad,Bernard:2006zz,Dalgic:2006gp,%
Laiho:2007pn}.
For recent reviews, including references to work in Asia, Australia, and
Europe, see~\cite{DellaMorte:2007ny,Juttner:2007sn}.

The past several months witnessed an unexpected development in flavor
physics, in the field of leptonic decays of pseudoscalar mesons.
(Pseudoscalar bosons have spin~0 and negative parity; the most common 
examples are the pion and kaon.)
One can consider the measured decay rates to determine a QCD matrix 
element called the decay constant, and denoted $f_\pi$ (for the pion),
$f_K$ (for the kaon), and so~on.
Table~\ref{tab:f} shows the results of some SciDAC-supported
calculations of decay constants and the corresponding measurements.
\begin{table}[tbp]
	\centering
	\caption[tab:f]{Decay constants $f_P$ of pseudoscalar mesons (here 
	$P\in\{\pi,K,D,D_s\}$).
	Values in MeV.
	Experimental determinations assume a suitable element of the CKM 
	matrix element consistent with all flavor data.}
	\label{tab:f}
	\vspace*{0.7em}
	\begin{tabular}{ccccc@{~}cc}
		\hline\hline
		Meson & MILC~\cite{Bernard:2007ps} & FNAL~\cite{Aubin:2005ar} & 
			HPQCD~\cite{Follana:2007uv} & Experiment & (Ref.) &  Deviation \\
		\hline
		$\pi$   & $128\pm3$  & -- & $132\pm2$ & 
			$130.7\pm0.4$ & \cite{Yao:2006px} & $0.4\sigma$  \\
		$K$     & $154\pm3$  & -- & $157\pm2$ & 
			$159.8\pm1.5$ & \cite{Yao:2006px} & $1.7\sigma$  \\
		$D^+$ & -- & $201\pm19$ & $207\pm4$ & 
			$206\pm9$ & \cite{Eisenstein:2008sq} & $0.1\sigma$ \\
		$D_s$ & -- & $249\pm17$ & $241\pm3$ & 
			$277\pm9$ & \cite{Dobrescu:2008er} & $3.8\sigma$ \\
		\hline\hline
	\end{tabular}
\end{table}
The calculations agree very well with the measurements, except in the
case of the charmed strange pseudoscalar meson~$D_s$.

This set of circumstances is odd, because the computation is easiest for
the $D_s$.
Algorithms for the propagators of light quarks slow down as the quark 
mass~$m_q$ is decreased.
As a consequence, properties of hadrons containing up or down quarks are 
reached via a (controlled) extrapolation in~$m_q$.
This introduces a source of uncertainty that is absent for the $D_s$,
which, as the lightest bound state of a charmed quark and a strange
antiquark, can be computed by working directly at the physical quark
masses.

There is no obvious explanation for this discrepancy within the Standard
Model.
Unless there is a real blunder somewhere, the discrepancy points to a 
non-Standard particle mediating the decays $D_s\to\mu\nu$ and 
$D_s\to\tau\nu$, thereby changing the interpretation of the 
``measured'' $f_{D_s}$.
It turns out to be rather easy to devise models of such
interactions~\cite{Dobrescu:2008er}.
Given the agreement of CLEO's new measurement of
$f_{D^+}$~\cite{Eisenstein:2008sq}, new $W'$ bosons and charged Higgs
bosons seem unlikely, making leptoquarks---particles with baryon and
lepton number---are the most plausible agents of decay.

\section{Nucleon structure and nuclear physics}
\label{sec:kern}

Let us turn now to nuclear physics, where the importance of lattice QCD 
is difficult to overstate.
Many nuclear laboratories, such as Jefferson Laboratory in the U.S.,
have turned their attention to the structure of the proton (the
simplest nucleus!)~\cite{Hagler:2007hu}, the excitation spectrum of
baryons~\cite{McNeile:2007fu}, and processes that may produce
glueballs.
On the theoretical side, the previous decade witnessed dramatic growth 
in the application of effective field theories to understand nuclear 
interactions~\cite{Kaplan:2006sv}.
These techniques connect a wealth of nuclear and hadronic data to the 
QCD Lagrangian.
Lattice QCD is an irreplaceable tool to interpret the new and upcoming
experiments at JLab and to compute the so-called low-energy constants of
the effective theory~\cite{Beane:2008dv}.

A topic at the interface between nuclear and particle physics concerns
the distributions of partons (a generic term for quarks and gluons)
inside the proton.
In high-energy collisions of $p\bar{p}$ (at the Tevatron) or $pp$ (at 
the LHC), the transverse motion of partons can be neglected.
To compute cross sections (to produce, say, a Higgs boson), the key
property needed is the distribution of the longitudinal momentum.
From a nuclear perspective, this is also one of the most basic aspects of 
proton structure.
Usually one studies parton distributions $q$ as a function of the 
fraction $x$ of longitudinal momentum carried by a parton, $0<x<1$.
Fits to scattering data can determine $q(x)$ only over a range,
$x_{\rm min}<x<x_{\rm max}$, limited by kinematics and statistics.
Lattice QCD can compute moments of $q$:
$\langle x^n\rangle_q =\int_0^1dx\,x^nq(x)$.
Clearly the two kinds of information are complementary.
The experimental range can be extended with \emph{ad hoc} functions
possessing the right asymptotic forms at the endpoints, and in this
way an Ansatz-guided experimental determination of 
$\langle x^n\rangle_q$ is possible.
One finds good agreement with lattice calculations~\cite{Hagler:2007xi}.
To explore nucleon structure further, the next step is to probe the 
transverse structure and compute the so-called generalized parton 
distributions~\cite{Hagler:2007xi}.

The axial charge of the nucleon is another topic of keen interest, 
because it appears in the description of $\beta$ decay.
It is also related to the relative spin polarization of up and down 
quarks in the proton.
Some~recent results with lattice QCD are
\begin{equation}
	g_A = \left\{ \begin{array}{rl}
		1.21 \pm 0.08 & \textrm{Ref.~\cite{Edwards:2005ym}} \\
		1.20 \pm 0.07 & \textrm{Ref.~\cite{Yamazaki:2008py}}
	\end{array} \right.,
\end{equation}
showing good agreement with experiment ($g_A=1.269\pm0.003$).
Work is under way to reduce the theoretical uncertainty.

In all fields of physics, spectroscopy is a time-honored approach to 
learn about the dynamics of the underlying phenomena of interest.
In nuclear and hadronic physics, therefore, an important area of 
investigation is the excitation spectrum of hadrons, particularly 
baryons.
Excited states present additional challenges for lattice QCD, but the
tools needed are available and have been demonstrated to
work~\cite{Sasaki:2001nf,Basak:2007kj}.
This subject extends to include topics such as the intriguing Roper
resonance~\cite{Mathur:2003zf} and electromagnetic transitions of the
form $\gamma N\to N^*$, where $N^*$ is an excited
nucleon~\cite{Lin:2008qv}.

Because the gluons couple to each other, it is conceivable that there 
are hadrons consisting essentially of gluons, with no valence quarks.
(Virtual quark-antiquark pairs appear in all hadrons.)
The most persuasive evidence that glueballs do indeed exist come 
from lattice-QCD calculations.
In an approximation omitting virtual quark pairs, the evidence is clear
that stable gluon-only bound states
arise~\cite{Vaccarino:1999ku,Morningstar:1999rf}.
There are two challenges with glueballs.
One, obviously, is to extend the lattice calculations to full QCD with 
virtual quark-antiquark pairs.
The other is to identify them unambiguously in the 
laboratory~\cite{Sexton:1995kd}.
Experiments with $\gamma p$ collisions are expected to produce such 
states and are planned at the 12~GeV upgrade of CEBAF, the 
accelerator at Jefferson Lab.
To plan this program it is essential to know the photocouplings of 
the mesons (exotic or not).
It is possible to compute these photocouplings in lattice QCD, as has 
been demonstrated for charmonium ($\bar{c}c$ mesons) \cite{Dudek:2006ej}.
An extension of this work to light mesons, including exotics and 
glueballs, is under~way.

A long-term goal of nuclear physics is to compute many-body 
interactions from QCD.
In addition to providing a better understanding of the nucleus, these 
calculations are relevant to mesonic atoms, namely, those 
containing a positively and negatively charged meson, and to 
strangeness in neutron stars.
Although in some cases lattice QCD results can be compared to 
measurements, the more interesting situation is where this cannot be 
done.
The aim here is to gain confidence in the reliability of these 
difficult calculations, so that one can use them (with robust 
uncertainty estimates) in applications where it is otherwise impossible 
to acquire the needed information any other way.

An interesting example is the lattice calculation of the $\pi K$ 
scattering lengths, of which there are two, $a_{1/2}$ and $a_{3/2}$, 
depending on whether the $\pi K$ system has isospin $I=\frac{1}{2}$ or 
$\frac{3}{2}$.
A~recent lattice-QCD calculation finds~\cite{Beane:2006gj}
\begin{eqnarray}
	a_{1/2} m_\pi c/\hbar & = & +0.1725 \pm 0.0017^{+0.0023}_{-0.0156},
	\label{eq:a1/2} \\
	a_{3/2} m_\pi c/\hbar & = & -0.0574 \pm 0.0016^{+0.0024}_{-0.0058},
	\label{eq:a3/2}
\end{eqnarray}
in convenient dimensionless units.
These are in rough, but not spectacular, agreement with other 
determinations using chiral perturbation theory or the Roy-Steiner 
equations.
For example, an analysis based on the Roy-Steiner
method~\cite{Buettiker:2003pp} finds
\begin{eqnarray}
	a_{1/2} m_\pi c/\hbar & = & +0.224  \pm 0.022,  \\
	a_{3/2} m_\pi c/\hbar & = & -0.0448 \pm 0.0077,
\end{eqnarray}
employing an extrapolation from high energies down to the threshold 
(where the scattering length is defined);
see~\cite{Beane:2008dv} for more discussion.
A more direct experimental determination is proposed by the DiRAC 
Collaboration, which would form $\pi^-K^+$ atoms~\cite{DiRAC}.
In this sense, Eqs.~(\ref{eq:a1/2}) and~(\ref{eq:a3/2}) represent
something exciting: a theoretical prediction awaiting definitive
experimental confirmation.

Multimeson systems are proving to be a fruitful area of research, with
further work on $\pi\pi$~\cite{Beane:2007xs}, $KK$~\cite{Beane:2007uh},
and even states with as many as twelve pions~\cite{Detmold:2008fn}.
Once methods have been vetted in mesonic systems, the next step is to 
consider nucleon-nucleon~\cite{Beane:2006mx} and
nucleon-hyperon~\cite{Beane:2006gf} systems.
This is more difficult, because the signal-to-noise ratio in the Monte 
Carlo estimate of baryonic correlation functions is worse.
Here the methods of effective field theories can be used to extend 
feasible lattice-QCD calculations to a wider range of nuclear phenomena.
The hyperon-nucleon interaction is of interest, because there is
speculation that the large Fermi energy may make it energetically
favorable for some nucleons to transmute to
hyperons~\cite{Kaplan:1986yq}.

\section{QCD thermodynamics}
\label{sec:thermo}

The discussion in the introduction of the computational scope of
numerical lattice QCD mentions that, but does not explain why, the
spacetime lattice usually has a time extent much longer than the spatial
dimension, for example, $32^3\times128$.
The reason is that the functional integral actually describes a grand 
canonical ensemble (in the sense of thermodynamics) at temperature~$T$,
\begin{equation}
	k_BT = \hbar c/N_ta,
	\label{eq:temperature}
\end{equation}
where $N_t$ is the number of sites in the time direction, $a$ is the 
lattice spacing, and $k_B$ is Boltzmann's constant.
In the applications to particle and nuclear physics discussed in 
Sections~\ref{sec:sm} and~\ref{sec:kern}, one wants $T=0$ and, hence, 
takes $N_t$ large enough for the Boltzmann suppression 
$e^{-mc^2/k_BT}=e^{-N_tamc/\hbar}$ to eliminate thermal effects, where 
$m$ is a hadron mass.

There are, however, many areas of research where a nonzero 
temperature and a nonzero chemical potential arise.
For nonzero temperature, one simply makes $N_t$ smaller, typically 
$4\leq N_t\leq 12$ in current work.
Such lattices take somewhat less memory, but not less computing because 
thermodynamics obviously requires calculations at several temperatures.
It is conceptually easy to introduce a nonzero chemical potential into 
lattice gauge theory, but the conceptually clean approach is 
computationally beyond present resources.
For small chemical potential, several approaches are available and 
efficient~\cite{Schmidt:2006us}.

\begin{figure}[tp]
	\begin{minipage}{0.48\textwidth}
		\includegraphics[width=204pt]{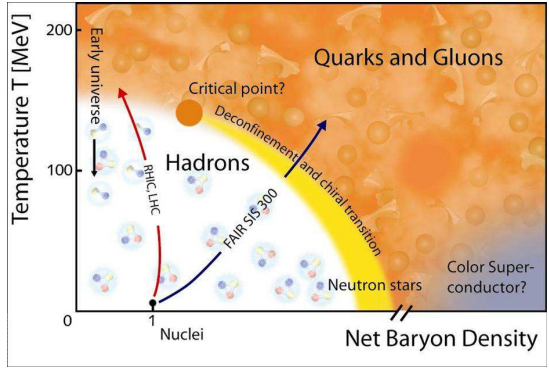}
	\end{minipage}
	\hfill
	\begin{minipage}{0.48\textwidth}
		\caption[fig:phase-diagram]{Phase diagram of QCD, showing
		conjectured phases and phenomena of physical interest.
		Accelerator experiments at RHIC and LHC probe small chemical 
		potential, around the crossover regime.
		At the planned Facility for Antiproton and Ion Research (FAIR)
		experiments can reach larger chemical potential.
		Image from the CBM Collaboration~\cite{cbm}.}
		\label{fig:phase-diagram}
	\end{minipage}
\end{figure}

The physical phenomena of interest are summarized by the phase diagram
in figure~\ref{fig:phase-diagram}.
The key feature is a phase transition from hadronic matter to a 
substance called the quark-gluon plasma.
The transition is expected to be first order and end at a critical point
at chemical potential $\mu\neq 0$.
Between the critical point and the $\mu=0$ axis the transition is a 
rapid crossover.
Lattice-QCD calculations, as well as measurements of heavy-ion 
collisions at RHIC and LHC, probe the crossover, which is relevant to 
the cooling of the early universe.
Neutron stars appear at high baryonic density, namely, high~$\mu$.
At even higher densities, several phases with color superconductivity 
are thought to exist.
Apart from heavy-ion collisions (so far at small~$\mu$) much of this 
phase diagram is unexplored, and lattice gauge theory offers the only 
\emph{ab initio}, quantitative, theoretical tool to see if it really 
reflects~QCD.

The central concept in QCD thermodynamics is the equation of state
(EoS), namely, the relationship between the pressure~$p$ and energy
density~$\epsilon$ as the temperature~$T$ and chemical potential~$\mu$
are varied.
The EoS reveals the relevant degrees of freedom.
In QCD, are they hadrons?  Or quarks and gluons?
At low temperatures, hadrons are clearly present, and phenomenological 
models of hadron gases can describe QCD thermodynamics.
At very high temperatures, perturbation theory in 
$\sqrt{\alpha_s(\hbar c/k_BT)}$ can be used.
But near the transition, nonperturbative methods are needed.

The first step is to find the transition temperature, which has now been
done with 2+1 flavors of realistic sea
quarks~\cite{Bernard:2004je,Cheng:2006qk}.
The different physics below, near, and above the critical temperature
$T_c$ also has ramifications for controling systematic errors.
Discretization effects from the lattice, for example, depend on the 
basic degrees of freedom.
(For a crossover it is, strictly speaking, incorrect to speak 
of a critical temperature.  But the critical point in the phase diagram 
is close enough to the $\mu=0$ axis that thermodynamic variables change 
rapidly at $T_c$.)

The next step is to map out the thermal contribution to the trace of 
the energy-momentum tensor, $\tr\Theta=\epsilon-3p$.
From the temperature dependence of $\epsilon-3p$ other thermodynamic 
variables, such as the entropy density, can be worked out.
This also has now been done with 2+1 flavors of realistic sea
quarks~\cite{Bernard:2006nj,Cheng:2007jq}, including at small 
$\mu$~\cite{Bernard:2007nm}.
As the system is heated, even to $2T_c$, the behavior of $\epsilon-3p$
shows that the system does not turn rapidly into weakly interacting 
quarks and gluons.

To make contact with heavy-ion collisions, the last step would be to 
take the output of lattice QCD as input to hydrodynamic models.
With these models one would then compute experimental observables, such 
as elliptic flow and quarkonium suppression.
With current computing resources, the EoS has been pinned down with 
20--25\% accuracy.
To make a big impact on hydrodynamic modeling, however, it will be 
necessary to reduce the error to 5\%.
This is conceptually straightforward to achieve, but it will be 
computationally challenging.
Part of the strategy is to reduce discretization effects, which scale
as $a^2$ ($a$ is the lattice spacing), while the net cost scales as
$a^{-11}$.

\section{Non-Standard particle physics}
\label{sec:bsm}

As mentioned above, particle physics awaits a new era when the LHC 
starts producing physics results.
Commencing later this year, the LHC will collide two beams
of 7~TeV (TeV = teraelectronvolt) protons.
(Amusingly, some particle physicists now call this energy scale the
terascale, deliberately alluding to terascale computing.
It would be unkind to tell them that in computing we are moving on to 
the petascale.)
In particular, the LHC will address the central problem of particle physics, 
which is the origin of the $W$ and $Z$ boson masses.
The fact that these masses do not vanish demonstrates that some 
mechanism breaks the electroweak gauge symmetry spontaneously.
``Spontaneous symmetry breaking'' means simply that, although the 
equations of motion are symmetric, the solution chosen is not symmetric.
For example, in the Standard Model a scalar field breaks electroweak 
symmetry, via a technique invented by Higgs.
Interactions between the Higgs field and the fermions also generate
(in the Standard Model) the quark and charged lepton masses and the
CKM matrix.
Therefore, flavor physics will remain interesting in the LHC~era, and
the physics program discussed in Section~\ref{sec:sm} will need to be 
continued and extended.

In this section, however, I would like to consider applications of 
lattice gauge theory to understanding the mechanism of electroweak 
symmetry breaking.
Since we suspect that the Standard Higgs sector is far from the full 
story, it is possible (some would say likely) that strongly coupled 
fields are involved.
For example, instead of being an elementary field, the Higgs boson (and
its siblings, the longitudinal polarization states of the $W$ and $Z$
bosons) could be composite.
The constituents would then form a whole spectrum of states, just as 
quarks bind into many different hadrons.
Even if the Higgs is elementary at the terascale, other agents of 
electroweak symmetry breaking could be strongly coupled.

Numerical work on theoretical bounds on the Higgs mass are a
long-standing topic in lattice field
theory~\cite{Dashen:1983ts,Heller:1993yv}.
The issue is that the Higgs mass is proportional to its self-coupling,
which runs at short distances in a way similar to $\alpha$ in QED;
see figure~\ref{fig:run}.
The perturbative running becomes invalid once the coupling is too 
large, and nonperturbative methods are required.
A~subtle relation arises between the Higgs mass and a scale where new 
phenomena imply that the Standard Model breaks down.
To make a long story short, one can in this way bound the Higgs mass 
from above.
A similar analysis applied to the Yukawa coupling between the Higgs 
field and the top quark leads to a lower bound on the Higgs mass, 
as discussed in a recent comprehensive analysis~\cite{Fodor:2007fn}.

A popular class of non-Standard models is to replace the Higgs sector 
with a strongly interacting sector that is, in some respects at 
least, analogous to QCD.
In fact, absent a stronger source of symmetry breaking, the pions of QCD
would break the electroweak symmetry, leading to $W$ and $Z$ masses
around 100~MeV.
What one needs, then, is a new quantum number, called technicolor, 
and a gauge force to bind the corresponding particles, called 
techniquarks, together.
These dynamics are posited to take place at terascale energies and, 
by analogy with QCD, generate vector boson masses around 100~GeV, as 
observed.
The earliest models of technicolor followed the example of QCD too 
closely and have been ruled out by experimental measurements.
One way around the experimental constraints is to assume that the
coupling evolves more slowly than in QCD.
This is called ``walking'' (i.e., slower than running) technicolor.
For a wide range of scales below the confining scale, the coupling is 
too strong for perturbation theory to be relied on, so 
nonperturbative work with lattice gauge theory is called for.
In the past year, the walking hypothesis has indeed been tested,
by examining the scaling behavior of the technipion mass, the vector 
meson mass, and the technipion decay constant~\cite{Catterall:2007yx}.
Even more recently, the evolution of $\alpha_s$ in (a TeV-scale version
of) QCD with several flavors has been computed~\cite{Appelquist:2007hu},
to see how many flavors are needed to having a walking coupling, and how
a walking coupling changes the physics.

The most popular way to relieve the theoretical problems of the
Standard Model is to introduce supersymmetry, a symmetry that
transforms fermions into bosons, and vice versa.
In these models, every known particle (quarks, leptons, gauge bosons) 
comes with another state known as a superpartner.
In the minimal supersymmetric model, the Higgs sector contains three 
neutral bosons, a charged pair, and their superpartners.
No superpartner has been observed, so at the terascale and below
supersymmetry must be broken somehow.
Usually supersymmetric models are conceived of as weakly coupled, but,
since they are gauge theories, strong coupling is not to be dismissed
out of hand.
It is therefore of great interest to formulate and simulate
supersymmetric gauge theories on a spacetime lattice.
This has been an extremely fruitful theoretical field during the past
several years~\cite{Kaplan:2003uh,Giedt:2007hz}, and now first
numerical simulations are  being carried
out~\cite{Catterall:2007fp,Catterall:2008yz}.
This rapid development yields hope that nonperturbative studies of 
supersymmetric models will be possible, including studies of 
supersymmetry breaking.

\section{Perspective}

Every now and then one hears the question, ``How much computing would it 
really take to solve QCD on a computer?''
The questioner is (usually) well meaning, but most experts would 
agree that the question is not well conceived.
The calculations for some of the simplest hadronic quantities are 
finally becoming mature, with analyses that forthrightly address all 
sources of uncertainty.
This class of observable, sometimes called
``gold-plated''~\cite{Davies:2003ik}, consists of masses of particles
that are stable (with respect to the strong interactions), not too close
to thresholds, and of hadronic transition matrix elements with one or
zero of these stable hadrons in the final and initial states.
Many, but not all, aspects flavor physics (Section~\ref{sec:sm}), and
some aspects of hadronic structure (Section~\ref{sec:kern}), lie in this
class, and the first solid round of serious computation is nearing 
its~end.

But this is the end of the beginning.
In these nearly mature fields, it turns out that precision is 
important, at least for the foreseeable future.
For example, to gain persuasive evidence of new phenomena in flavor 
physics, 5--25\% accuracy is not enough.
A better target (given the stakes and the quality of the 
corresponding experiments) is 1--3\%.
If the history of kaon physics is a guide, it will eventually be 
important to go further.
Happily, the need for and clear path to precision has persuaded the
particle-physics community that the resources needed for lattice 
flavor physics are worth the cost.

Even a quick glance at Section~\ref{sec:kern} reveals that the large
majority of interesting problems in nuclear physics require much more 
computing resources than gold-plated observables.
By definition the excited hadrons are not stable under the strong 
interaction.
They will be harder to pin down, and it is possible that we will not 
know how hard until the work currently under way reaches some of its 
milestones.
The calculations of scattering lengths, while impressive, must be 
repeated with several values of the physical volume to ensure that 
the mathematical formalism used to extract them works as it should.
Given the importance of nuclear lattice QCD in guiding nuclear
experimentation, one would hope for full support from experimenters
for the needed computing resources.

The ``how much?''~question should be rephrased.
It is really a class of questions, ``How much computing would it take 
to compute my favorite observable?''
Of course, the answer depends on your observable and its wider context.
It should be clear, however, that physics with either precision and
complexity leads to questions whose answers will need one or two orders
of magnitude more computing than what is available now.

Another reason the ``how much?''~question is ill-conceived is that it 
often overlooks the heterogeneous job mix needed to attack many 
problems in lattice QCD.
It is not enough to generate huge ensembles of lattice gauge fields, 
with ever smaller lattice spacing and ever smaller quark masses, on 
the world's highest-capability computer.
This step is necessary, creating a gold mine of information.
It is not enough, however, merely to open a mine; the gold must be
extracted and, here, we need a large capacity of computing that is
flexible in every way (while having significant capability of its own).

We do not know how broad the future of lattice gauge theory
will be.
If the LHC discovers a strongly interacting sector beyond the 
Standard Model, it is likely that chiral fermions play a role.
This is simply because, at the terascale, we already know that 
left-handed fermions and right-handed fermions are different fields.
But chiral fermions possess some conceptual and computational 
challenges not found in QCD.
A~strongly coupled terascale will bring many more particle theorists to
lattice gauge theory, and this larger community will clamor for
computing that is up to the task.

\ack

This work has been supported in part by the United States National Science
Foundation and the Office of Science of the U.S.\ Department of Energy
(DOE).
Software development and hardware prototyping within the USQCD 
Collaboration are supported by SciDAC.
Fermilab is operated by Fermi Research Alliance, LLC, under Contract
DE-AC02-07CH11359 with the US~DOE.

\providecommand{\newblock}{}

\end{document}